\documentclass[reprint,aps,prd,showpacs,showkeys]{revtex4-1}

\usepackage{natbib}
\usepackage{graphics}
\usepackage{bm}
\usepackage{hyperref}
\usepackage{amsmath,amssymb}

\begin{document}

\title{Plasmons in QED Vacuum}

\author{E.~Yu.~Petrov}
\author{A.~V.~Kudrin}
\email{kud@rf.unn.ru}
\affiliation{Department of Radiophysics, University of Nizhny Novgorod,
23 Gagarin Ave., Nizhny Novgorod 603950, Russia}
\begin{abstract}
The problem of longitudinal oscillations of an electric field and
a charge polarization density in QED vacuum is considered.
Within the framework of semiclassical analysis, we calculate
time-periodic solutions of bosonized (1+1)-dimensional QED (massive
Schwinger model). Applying the Bohr--Sommerfeld quantization
condition, we determine the mass spectrum of charge-zero bound
states (plasmons) which correspond in quantum theory to the found classical
solutions. We show that the existence of
such plasmons does not contradict any fundamental physical
laws and study qualitatively their
excitation in (3+1)-dimensional real world.
\end{abstract}


\maketitle

\section{INTRODUCTION}

It is a common knowledge that homogeneous plane electromagnetic waves in vacuum are transverse.
The classical Maxwell equations in free space do not admit solutions which
correspond to the longitudinal waves, i.e., waves with the electric-
or magnetic-field vector collinear to the wave vector. Despite this well-known
fact, in recent years a substantial degree of interest has been shown in
various theoretical generalizations of Maxwell electrodynamics, which
incorporate longitudinal waves, and in an experimental search for such waves
in the radio frequency band~\cite{VVl,Khv,Pod,Mon}. However, it is clear
that within the framework of the purely classical electrodynamics, any
theoretical constructions of this type will lead to violation of
the fundamental physical principles such as Lorentz invariance and
the charge conservation law.

The situation changes in quantum electrodynamics (QED)
due to virtual electron--positron pairs and vacuum polarization. Considering
QED vacuum as a plasma of virtual electrons and positrons, one can suppose
that longitudinal (Langmuir) oscillations should exist in such a medium.
Continuing the analogy between the usual plasma and polarized vacuum, we can
intuitively estimate possible characteristic frequencies of longitudinal
oscillations of the electric field (plasmons) in QED as the frequencies
comparable with the electron Compton frequency.

The nomenclature commonly accepted in QED refers to the fields described by the
irrotational component of the four-potential as
longitudinal waves~\cite{Dir,Fey1,Fey2}.
It is known~\cite{Fey1}, that such fields are nonphysical. However,
this circumstance does not mean that longitudinal oscillations
of an electric field cannot exist in principle.
The problem of existence of longitudinal electric fields
has direct bearing on the fundamental QED issue
of an infinite electron self-energy, which is due to
the instantaneous
Coulomb interaction approach~\cite{Fey2}.
It is well known, that QED is a local theory in the sense that it considers interaction of point particles with an electromagnetic field, and the action within the framework of this theory is local. However, unlike the classical point electron, the Dirac electron possesses internal degrees of freedom, which are specified by the Dirac matrices. This leads to the quantum nonlocality effect discussed in, e.g.,~\cite{Efr1,Efr2}. Allowance for such nonlocality on the Compton scales makes it possible to resolve the fundamental QED issue of an infinite electron self-energy. Obviously, the Coulomb law cannot be applicable at small distances of the order of the electron Compton wavelength. The Coulomb field, as a solution of the Maxwell equations, cannot satisfy exactly nonlinear QED equations of coupled electromagnetic and spinor fields. The influence of the nonlocality effects can be taken into account via modification of constitutive relations
in Maxwell electrodynamics (effective nonlocal field theory) and, as will be shown below, leads to the possibility of longitudinal modes. The problem is thus related to the issues of causality and nonlocality in quantum theory, and is far from trivial.

Meanwhile, there is currently a great deal of interest in the
dispersive vacuum effects~\cite{Roz,Gri,Gus,Dit,Dun,Mar,Lun}. Recent advances in the laser
technology, such as the construction of powerful X-ray free electron lasers, make
the fundamental effects related to the dispersive properties of quantum
vacuum achievable in laboratory experiments~\cite{Alk,Rob,Dip1,Ruf,Dip2,Boh}.
Thus, the poorly studied problem of existence of plasmons in QED vacuum is of
not only theoretical, but also practical importance.

The well-known method of solving some complicated problems and obtaining
nonperturbative results in quantum field theory suggests to treat the
Heisenberg operator field equations as $c$-number field equations and find
classical solutions to them. Some quantum properties can then be retrieved
through semiclassical methods~\cite{Das1,Das2,Jac}. We cannot apply
this approach directly to the general QED equations of coupled
electromagnetic and fermionic fields. While classical solutions of
the purely Bose theory correspond to the coherent states of bosons,
the classical limit of the Fermi fields, due to the exclusion principle
and anticommutation relations, leads to some abstract construction
with Grassmann numbers, the physical interpretation of which is challenging to obtain.
However, the
above-mentioned method
can be applied to our problem if we limit ourselves to consideration
only of one-space and one-time dimensions. In this case, we can bosonize
QED, find time-periodic classical solutions, and then quantize these
solutions using the Bohr--Sommerfeld quantization rule.

In this article, we present a simple semiclassical analysis of plasmons in
quantum vacuum on the basis of bosonized QED in 1+1 dimensions, i.e., the massive
Schwinger model~\cite{Sch,Col1,Col2,Fis,Ada,Sad,Byr,Nag,Nan,Kle}. We describe
a technique for calculating the spectra of plasmons and perform some qualitative
study of their excitation in (3+1)-dimensional real world.

\section{BASIC EQUATIONS}

The Lagrangian density of QED in 1+1 dimensions is given by~\cite{Sch,Col1,Col2,Fis}
\begin{eqnarray}
{\cal L}=&&i\bar{\psi}\gamma^{\mu}\partial_{\mu}\psi-g\bar{\psi}\gamma^{\mu}\psi A_{\mu}
-m\bar{\psi}\psi\nonumber\\
&&
-{1\over 4}F^{\mu\nu}F_{\mu\nu}-{1\over 2}(\partial_{\mu}A^{\mu})^2.\label{eq1}
\end{eqnarray}
The rationalized (Lorentz--Heaviside) natural units with $\hbar=c=1$ are employed
and the following standard notations are used throughout: $\mu,\nu=0,1$, $x^{0}\equiv t$,
$x^{1}\equiv x$, $\psi$ is a two-component spinor field, and the $\gamma$ matrices obey
the relation $\gamma^{\mu}\gamma^{\nu}+\gamma^{\nu}\gamma^{\mu}=2{\rm g}^{\mu\nu}$ with
the metric tensor ${\rm g}^{\mu\nu}={\rm diag}(1,-1)$. The Levi-Civita symbol
is defined as $\varepsilon^{01}=-\varepsilon^{10}=1$. The electromagnetic field
tensor $F_{\mu\nu}=\partial_{\mu}A_{\nu}-\partial_{\nu}A_{\mu}$, $g$ is a coupling
constant with the dimension of mass, and $m$ is the bare mass of the electron.

The theory can be mapped into an equivalent Bose form via the bosonization
rules~\cite{Col1,Col2,Fis,Col3,Man,Pog}
\begin{eqnarray}
{\cal N}_{m}[i\bar{\psi}\gamma^{\mu}\partial_{\mu}\psi]\rightarrow &&
{1\over 2} {\cal N}_{\Lambda}(\partial_{\mu}\phi)^2,\nonumber\\
{\cal N}_{m}[\bar{\psi}\gamma^{\mu}\psi]\rightarrow &&
{1\over \sqrt{\pi}}\varepsilon^{\mu\nu}\partial_{\nu}\phi,\nonumber\\
{\cal N}_{m}[\bar{\psi}\psi]\rightarrow &&
-{\exp(\gamma)\over 2\pi} {\Lambda}\, {\cal N}_{\Lambda}[\cos(2\sqrt{\pi}\phi)],
\label{eq2}
\end{eqnarray}
where $\Lambda=g/\sqrt{\pi}$, ${\cal N}_{\eta}$ denotes normal ordering with
respect to the mass $\eta$~\cite{Col2}, and $\gamma=0.577...$ is Euler's constant.
Applying relations~(\ref{eq2}), we will further consider the
Bose fields $\phi$ and $A^{\mu}$ as $c$-number quantities. The classical limit
of the bosonized version of Eq.~(\ref{eq1}) has the form
\begin{eqnarray}
{\cal L}=&&{1\over 2}(\partial_{\mu}\phi)^2-\Lambda\,\varepsilon^{\mu\nu}A_{\mu}\,\partial_{\nu}\phi+
{a\over 2\pi}g\Lambda\,\cos(2\sqrt{\pi}\phi)\nonumber\\
&&
+{1\over 2}(F_{01})^2-{1\over 2}(\partial_{\mu}A^{\mu})^2,\label{eq3}
\end{eqnarray}
where $a=m\,\exp(\gamma)/g$.
The field equations following from Lagrangian density~(\ref{eq3}) read
\begin{eqnarray}
&&
\square\phi+\Lambda E+a \pi^{-1}g^{2}\sin(2\sqrt{\pi}\phi)=0,
\label{eq4}\\
&&
\square A^{\mu}=j^{\mu},\label{eq5}
\end{eqnarray}
where $\square=\partial^{\mu}\partial_{\mu}$, $E=F_{01}$ is the electric field, and
$j^{\mu}=\Lambda \,\varepsilon^{\mu\nu}\partial_{\nu}\phi$ is the polarization
current. It is obvious from Eq.~(\ref{eq5}) that imposing the Lorentz gauge
condition
\begin{equation}
\partial_{\mu}A^{\mu}=0\label{eq6}
\end{equation}
ensures the fulfilment of the charge conservation law $\partial_{\mu}j^{\mu}=0$.
Constraint~(\ref{eq6}) also implies that $A^{\mu}$ can be written as $A^{\mu}=\varepsilon^{\mu\nu}\partial_{\nu}u$.
Thus, from Eq.~(\ref{eq5}) we have
\begin{equation}
\partial_{\nu}(\square u-\Lambda \phi)=0.\label{eq7}
\end{equation}
Integrating Eq.~(\ref{eq7}) yields $\phi=\Lambda^{-1}\square u$ and
\begin{equation}
E=\square u=\Lambda \phi.\label{eq8}
\end{equation}
We restrict ourselves to consideration only of the case where
the constant of integration for Eq.~(\ref{eq7}), known as the $\theta$ angle of the theory~\cite{Col2}, is zero. The case $\theta\neq 0$ corresponds to the
appearance of a background electrostatic field~\cite{Col2} and can also be of some physical interest. However, it should be noted that even a very small value of $\theta$ (in the natural units used in our work) corresponds actually to a giant (from the engineering viewpoint) electrostatic field. Thus, the choice $\theta=0$ seems more physically justified in the context of the studied problem.

Inserting Eq.~(\ref{eq8}) into Eq.~(\ref{eq4}) and introducing the dimensionless
variables $\xi=\Lambda x$ and $\tau=\Lambda t$, we obtain
\begin{equation}
\partial^{2}_{\tau}\phi-\partial^{2}_{\xi}\phi+\phi+a\sin(2\sqrt{\pi}\phi)=0.\label{eq9}
\end{equation}
This is the massive sine-Gordon (MSG) equation~\cite{Fis,Sad}, i.e.,
bosonized version of (1+1)-dimensional QED.

It is the main purpose of the forthcoming analysis to find physically
meaningful solutions of Eq.~(\ref{eq9}). In all the subsequent calculations,
for modeling the polarization of electron--positron vacuum we adopt (see, e.g.,~\cite{Heb})
\begin{equation}
g/m=0.303\approx \sqrt{4\pi\alpha},\label{eq10}
\end{equation}
where $\alpha$ is the fine-structure constant ($a\approx 5.88$ in this case).

The potential energy for the scalar field $\phi$ is given by
\begin{equation}
V(\phi)={1\over 2}\phi^{2}-{a\over 2\sqrt{\pi}}\cos(2\sqrt{\pi}\phi).\label{eq11}
\end{equation}
The dependence $V(\phi)$ in the case~(\ref{eq10}) is shown in Fig.\,1(a). It is
seen that $V(\phi)$ possesses four stable classical minima. However, only
one state $\phi=0$ is the true minimum corresponding to the stable state
in quantum theory. The other three minima will be unstable when tunneling effects
are taken into account~\cite{Vol,Col4}.

\begin{figure}[h]
\includegraphics{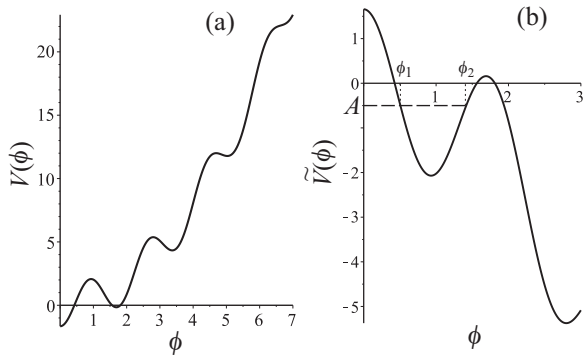}
\caption{(a) The potential for the MSG equation~(\ref{eq9}) in
the case~(\ref{eq10}) and (b) the inverted potential ${\tilde V}=-V$.}
\end{figure}

Let us consider briefly traveling wave solutions of Eq.~(\ref{eq9}),
which are apparently the unique case where exact analytical results can be
obtained. Substituting $\phi(\xi,\tau)=\phi(\eta)$, where $\eta=\xi-\beta\tau$ ($\beta<1$),
into Eq.~(\ref{eq9}), after simple algebra we obtain
\begin{equation}
\left({d\phi\over d\eta}\right)^{2}={2\over 1-\beta^2}[A+V(\phi)],\label{eq12}
\end{equation}
where $A$ is an integration constant. This equation formally corresponds to
the one-dimensional mechanical motion of a particle in an inverted potential ${\tilde V}=-V$.
The oscillations between two values $\phi_1$ and $\phi_2$, where
$\phi_1$ and $\phi_2$ are roots of the equation $A+V(\phi)=0$,
are possible if $A+V(\phi)>0$ for $\phi_1<\phi<\phi_2$ [see Fig.\,1(b)]. Thus,
the MSG equation~(\ref{eq9}) admits periodic traveling wave solutions.
The dispersion relation comprising the wavelength $\lambda$, the phase velocity $\beta$,
and the amplitude $A$ is given by
\begin{equation}
\lambda=\sqrt{2}\sqrt{1-\beta^{2}}\int^{\phi_2}_{\phi_1}\frac{d\phi}
{\sqrt{A+V(\phi)}}.\label{eq13}
\end{equation}
However, it can be shown that these solutions are modulationally unstable
even in classical theory~\cite{Whi}. The instability will also occur in
quantum theory due to the tunneling from one local minimum to another [see Fig.\,1(b)].
In view of the above, we will not dwell on this case.

The problem of finding other solutions can be solved only with
the help of numerical methods.

\section{STANDING WAVES AND SEMICLASSICAL QUANTIZATION}

We will seek time-periodic solutions of the MSG equation~(\ref{eq9}) on the
interval $|\xi|\le L/2$. Let the solutions satisfy the boundary
conditions
\begin{equation}
\phi(\xi=L/2,\tau)=\phi(\xi=-L/2,\tau)=0\label{eq14}
\end{equation}
and the time-periodicity condition
\begin{equation}
\phi(\xi,\tau+T)=\phi(\xi,\tau),\label{eq15}
\end{equation}
where $T=2\pi/\Omega$. The dimensional quantities $l$ and $\omega$, which correspond
to conditions~(\ref{eq14}) and~(\ref{eq15}), are defined by $x/l=\xi/L$ and
$\omega t =\Omega \tau$, respectively. Hence, $l\approx 5.85\,L/m$ and
$\omega\approx 0.17\,\Omega\,m$.

It follows from Gauss' law
\begin{equation}
\int^{L/2}_{-L/2}{\partial E\over\partial \xi}\,d\xi=\Lambda(\phi(L/2)-
\phi(-L/2))=Q=0\label{eq16}
\end{equation}
that conditions~(\ref{eq14}) imply the zero total charge.
Accordingly, the polarization current $j^{1}=-\partial_{t}E=-\Lambda\partial_{t}\phi$
vanishes at $\xi=\pm L/2$.

Before proceeding to numerical computations, let us give some elementary
consideration concerning a weak field limit. The simplest solution of the
linearized MSG equation~(\ref{eq9}), which satisfies
conditions~(\ref{eq14}) and~(\ref{eq15}), i.e.,
the lowest linear normal mode, is given by
\begin{equation}
\phi(\xi,\tau)=A\,\cos(\pi\xi/L)\,\cos(\Omega\tau),\label{eq17}
\end{equation}
where
\begin{equation}
\Omega^{2}=\Omega^{2}_{0}=1+2a\sqrt{\pi}+\pi^{2}/L^{2}.\label{eq18}
\end{equation}
Observe that the minimum frequency of the linear oscillations
is $\Omega_{0}(L=\infty)=(1+2a\sqrt{\pi})^{1/2}\approx 4.67$ ($\omega_{0}\approx 0.8\,m$).
Let us suppose that solution~(\ref{eq17}) survives in the weakly nonlinear case.
Taking into account the term proportional to $\phi^3$ in the series expansion
of $\sin(2\sqrt{\pi}\phi)$, one can find an approximate amplitude correction
to the dispersion relation~(\ref{eq18}):
\begin{equation}
\Omega^{2}=\Omega^{2}_{0}(L)-3a\pi^{3/2}A^{2}/4.\label{eq19}
\end{equation}
Equations~(\ref{eq17})--(\ref{eq19}) are useful for further analysis.

Any smooth solution of the boundary value problem
that is specified by Eqs.~(\ref{eq9}) and~(\ref{eq14})
and periodic in time with the period $T$ has a Fourier
representation. In our numerical calculations, we employ
the Fourier pseudospectral algorithm~\cite{Boy1,Boy2}. It is convenient
to seek the solution as the truncated Fourier series:
\begin{eqnarray}
\phi(\xi,\tau)=&&\sum\limits^{M}_{m=1}\sum\limits^{N}_{n=1}
\phi_{mn}\cos[(2m-1)\pi\xi/L]\nonumber\\
&&
\times\cos[(2n-1)\Omega\tau].\label{eq20}
\end{eqnarray}
Due to the odd nonlinearity, no even harmonics will appear. In order
to apply the iterative method, we write
\begin{equation}
\phi^{(i+1)}(\xi,\tau)=\phi^{(i)}(\xi,\tau)+\delta^{(i)}(\xi,\tau).\label{eq21}
\end{equation}
After linearization of Eq.~(\ref{eq9}) about the $i$th iterate $\phi^{(i)}$,
we obtain
\begin{eqnarray}
&&[\partial^{2}_{\tau}-\partial^{2}_{\xi}+1+2a\sqrt{\pi}
\cos(2\sqrt{\pi}\phi^{(i)})]\delta^{(i)}\nonumber\\
&&=\partial^{2}_{\xi}\phi^{(i)}-\partial^{2}_{\tau}\phi^{(i)}-\phi^{(i)}-a
\sin(2\sqrt{\pi}\phi^{(i)}).\label{eq22}
\end{eqnarray}
This equation is solved by expanding both $\phi^{(i)}$ and $\delta^{(i)}$ as double
Fourier series similar to Eq.~(\ref{eq20}) and demanding that the left- and right-hand sides
of Eq.~(\ref{eq22}) agree exactly at the $M\times N$ collocation points $\xi_k$ and
$\tau_j$ such that $0<\xi_{k}<L/2$ and $0<\tau_{j}<T/2$, where $k=1, \ldots , M$ and $j=1, \ldots , N$.
Equations~(\ref{eq21}) and~(\ref{eq22}) are iterated until $\delta^{(i)}$ is
negligibly small. At each step, one has to find the Fourier coefficients $\delta^{(i)}_{mn}$
from the linear system of $M\times N$ equations~(\ref{eq22}).

The method has fast convergence and explicitly reveals periodicity of
the desired solution. By means of this algorithm, we have built relatively
simple and fast FORTRAN code. The possible computational problems
inherent in the method are discussed in details in~\cite{Boy1}. First,
we must specify a sufficiently good guess for $\phi(\xi,\tau)$. Second,
if the parameters $L$ and $\Omega$ in Eq.~(\ref{eq20}) are varied, the matrix
of the linear system for the unknowns $\delta^{(i)}_{mn}$ will also vary, and
one or more matrix eigenvalues may cross zero.

Our numerical strategy for computing periodic solutions is the following.
We can fix the value of $L$ (say, $L=1$) and specify solution~(\ref{eq17}) with
$A=1$ as the guess (i.e., $\phi^{(1)}_{11}=A$, while all the other coefficients
$\phi^{(1)}_{mn}$ are equal to zero). In this case, setting $\Omega=\Omega_{0}(L)$
ensures that the iterations converge to the trivial solution, since Eq.~(\ref{eq17})
corresponds to the infinitesimally small values of $\phi$. Bearing Eq.~(\ref{eq19})
in mind
and taking $\Omega\lesssim \Omega_{0}$, one can compute the Fourier coefficients
for a finite-amplitude periodic solution. In all the computations, we use
$M=N=15$, which gives high accuracy. It turns out that a small random shift
of the collocation points from the nodes of the uniform grid improves convergence.

It is important that the correctness and possible computational errors
of the solution can easily be checked with the help of standard
mathematical software packages. Once the Fourier coefficients
$\phi_{mn}$ in Eq.~(\ref{eq20}) have been obtained, we can determine
what initial data $\varphi(\xi)=\phi(\xi,\tau=0)$ correspond to the found
periodic solution. Imposing the initial conditions $\phi(\xi,0)=\varphi(\xi)$
and $\partial_{\tau}\phi(\xi,0)=0$ along with the boundary conditions~(\ref{eq14}),
one can numerically integrate Eq.~(\ref{eq9}). We utilized the MAPLE intrinsic
solver ``pdsolve'' for this purpose.

\begin{figure}[h]
\includegraphics{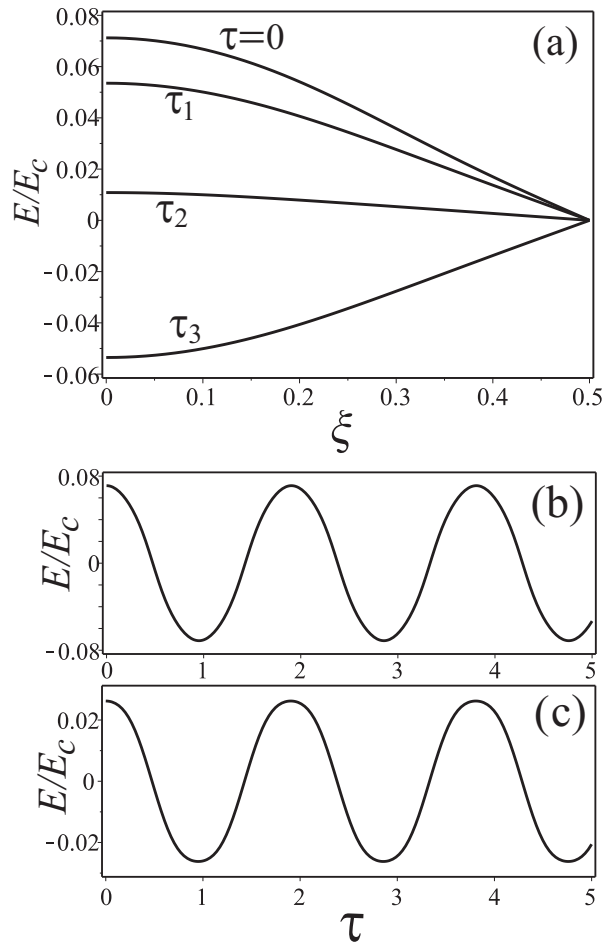}
\caption{(a) Normalized
electric field as a function of $\xi$ for $L=1$ ($l=5.85/m$) and $\Omega=3.3$ ($\omega=0.56\,m$)
at time instants $\tau=0$, $\tau_{1}=\pi/(4\Omega)$, $\tau_{2}=0.46\pi/\Omega$,
and $\tau_{3}=5\pi/(4\Omega)$.
Oscillograms of the field at (b) $\xi=0$ and (c) $\xi=0.35$.}
\end{figure}

Figure 2(a) shows the snapshots of the normalized field
\begin{equation}
E/E_{c}=g^{2}\phi/(m^{2}\sqrt{\pi})\label{eq23}
\end{equation}
at fixed time instants for $L=1$ ($l=5.85/m$) and $\Omega=3.3$ ($\omega=0.56\,m$).
Hereafter, $E_{c}=m^{2}/g$ is the critical Schwinger field strength.
Figures 2(b) and 2(c) show the oscillograms of the field at the points
$\xi=0$ and $\xi=0.35$, respectively. Although this case corresponds to
the strong nonlinearity (the amplitude of the fundamental is $\phi_{11}=1.352$),
the amplitudes of the higher harmonics are negligibly small (for example,
$\phi_{12}=-0.053$ and $\phi_{21}=0.074$) and the oscillations turn out
to be very close to monochromatic ones.

Fixing the quantity $L$ and consistently decreasing $\Omega$ from $\Omega_{0}(L)$ to
smaller values, one can obtain an amplitude-frequency characteristic which is the
dependence of the maximum field amplitude $E_{m}=E(0,0)$ of the time-periodic solution
on the fundamental frequency $\Omega$. Figure 3 shows the
amplitude-frequency characteristics for different values of $L$. It follows
from the performed computations that for $L\lesssim 1.9$ ($l\lesssim 11/m$), the
amplitude-frequency characteristics consist of four branches. Although the
standing wave solutions have no direct analogy with a one-dimensional motion,
the maximum field amplitudes for these branches approximately correspond
to the minima of the potential $V(\phi)$ [see Fig.\,1(a)].

Periodic solutions which correspond to the same $\Omega$ but different
maximum amplitudes (different branches) can be calculated numerically
by specifying different guess amplitude values. One should also
note that once some solution (i.e., the set of $\phi_{mn}$) has been obtained,
it can be used further as the guess itself.

The branches of the amplitude-frequency characteristics are connected
at the bifurcation points, where $dE_{m}/d\omega=\infty$. At these
points, the solution is not unique and iterative equation~(\ref{eq22})
is not soluble, since one or more matrix eigenvalues vanish~\cite{Boy1}.

\begin{figure}[h]
\includegraphics{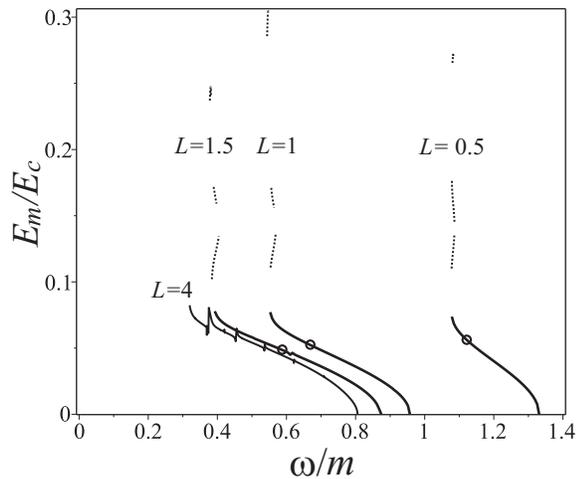}
\caption{Amplitude-frequency characteristics for different values of $L$.
The solution branches which correspond to the same $L$ and $\Omega$ but larger
field amplitudes are shown by the dotted curves. The circles on the solid
branches correspond to the values of
$\Omega$ for which Eq.~(\ref{eq26}) is satisfied with $k=1$.}
\end{figure}

Figure 4(a) illustrates the mode shape which corresponds to the same
parameters $L=1$ and $\Omega=3.3$ as for Fig.\,2, but for the second
branch in Fig.\,3. The field oscillograms of this solution are
presented in Figs.\,4(b) and 4(c). It is seen in Fig.\,4 that the
nonlinear effects become more pronounced here than those for
the first solution branch [see Fig.\,2]. The main Fourier
coefficients are $\phi_{11}=2.11$, $\phi_{12}=0.002$, and
$\phi_{21}=0.262$. Note that periodic oscillations
corresponding to the third and fourth solution branches,
despite their largest
amplitudes, are again close to monochromatic ones.

\begin{figure}[h]
\includegraphics{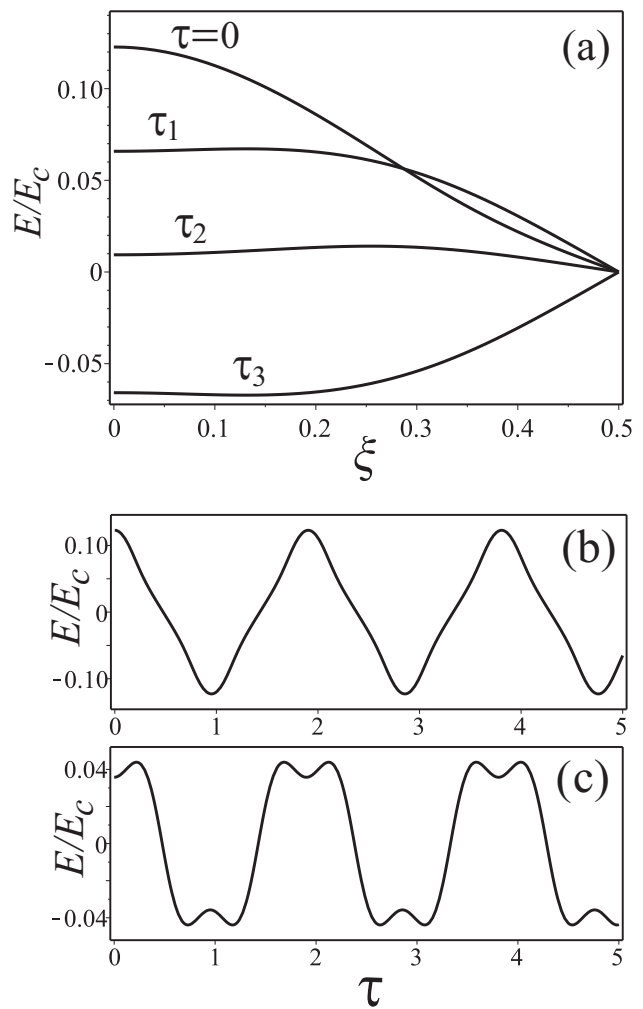}
\caption{The same as in Fig.\,2, but for
the next solution branch.}
\end{figure}

The above-described picture of periodic solutions changes for $L\gtrsim 2$.
For such values of $L$, we have found only single-branch amplitude-frequency
characteristics. An example of such a characteristic is shown in Fig.\,3
for $L=4$ ($l=23.4/m$). We can see that the characteristic for $L=4$
differs from other characteristics by the presence of small-amplitude
spikes. The peaks of the spikes correspond to the singular points
at which $dE_{m}/d\omega=\infty$. The appearance of these
spikes is due to the resonances of the higher
harmonics of the fundamental frequency inside our nonlinear ``cavity resonator''
with the ``mirrors'' at $\xi=\pm L/2$. Figures 5 and 6 illustrate how
the mode shape changes when the fundamental frequency switches from
$\Omega=2.3$ ($\omega=0.39\,m$) to $\Omega=2.2$ ($\omega=0.37\,m$)
which is closer to the resonance. With increasing $L$, the number of the
resonances increases, and overlapping of the resonances destroys
a periodic solution. Our calculations show that for $L\gtrsim 5$, the
existence of periodic solutions is a rare event.

\begin{figure}[h]
\includegraphics{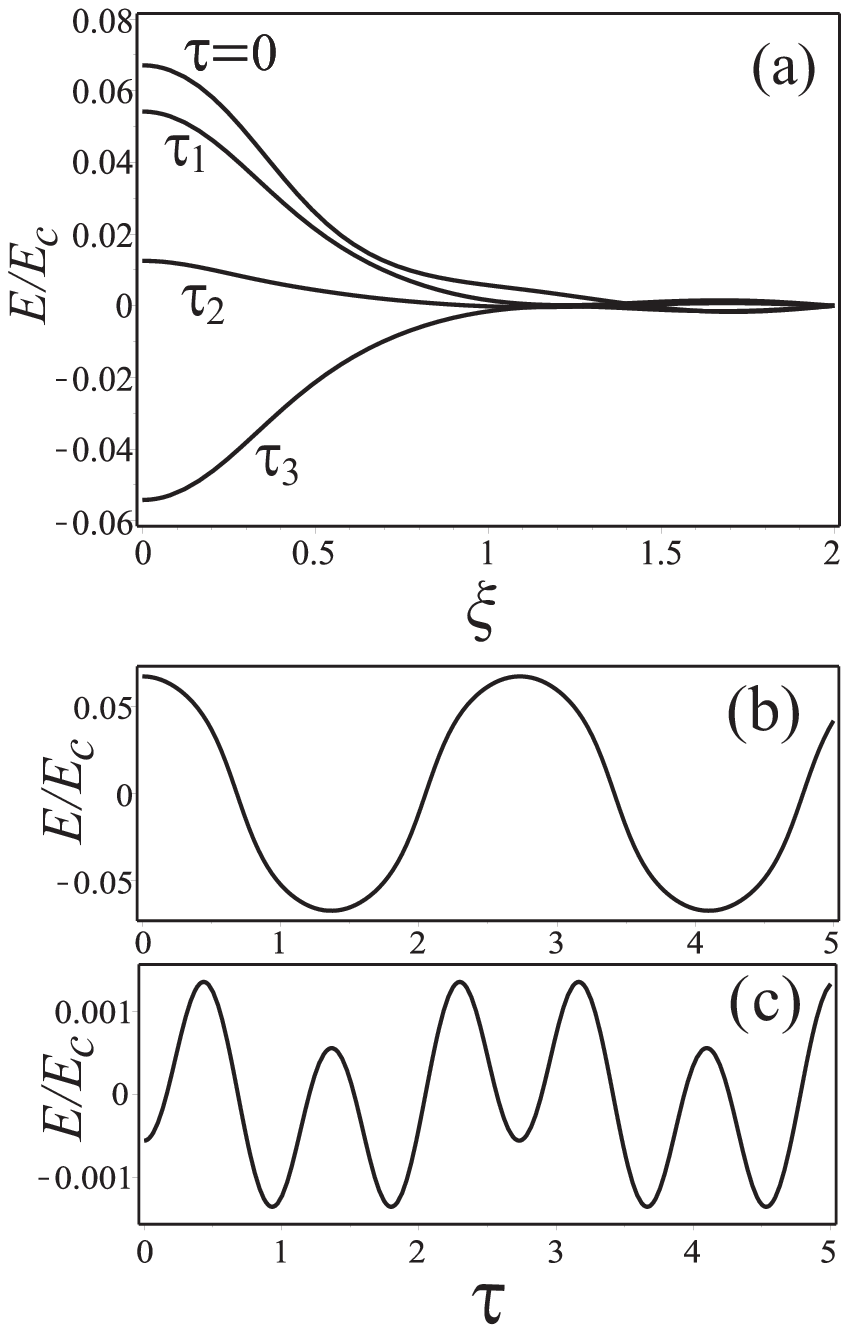}
\caption{(a) Normalized
electric field as a function of $\xi$ for $L=4$ ($l=23.4/m$) and $\Omega=2.3$ ($\omega=0.39\,m$)
at time instants $\tau=0$, $\tau_{1}=\pi/(4\Omega)$, $\tau_{2}=0.46\pi/\Omega$,
and $\tau_{3}=5\pi/(4\Omega)$.
Oscillograms of the field at (b) $\xi=0$ and (c) $\xi=1.5$.}
\end{figure}
\begin{figure}[h]
\includegraphics{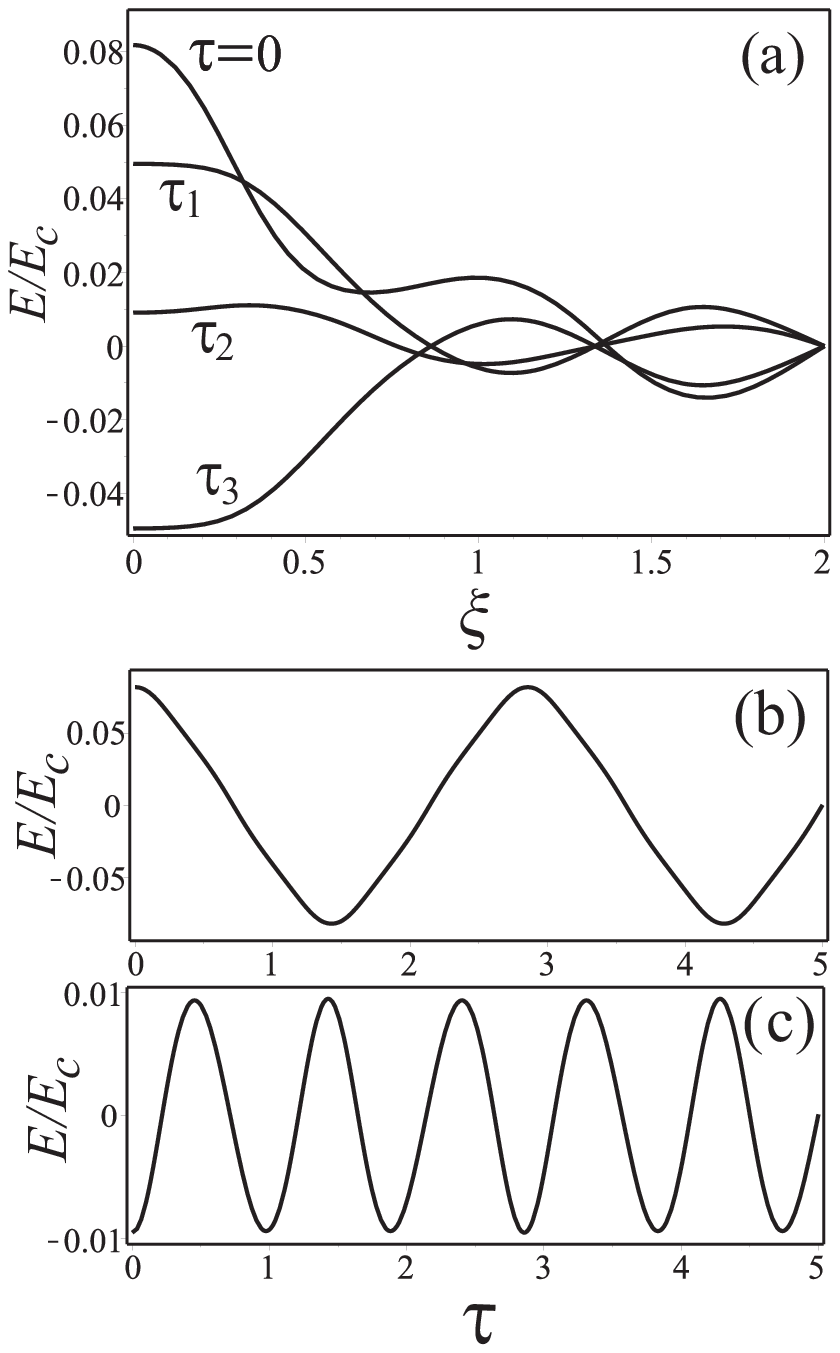}
\caption{(a) Normalized
electric field as a function of $\xi$ for $L=4$ ($l=23.4/m$) and $\Omega=2.2$ ($\omega=0.37\,m$)
at time instants $\tau=0$, $\tau_{1}=\pi/(4\Omega)$, $\tau_{2}=0.46\pi/\Omega$,
and $\tau_{3}=5\pi/(4\Omega)$.
Oscillograms of the field at (b) $\xi=0$ and (c) $\xi=1.5$.}
\end{figure}

It is interesting to consider briefly some results of numerical
integration for large $L$. Figure 7 illustrates two oscillograms
of the solutions $\phi_{1}(\xi,\tau)$ and $\phi_{2}(\xi,\tau)$
of the boundary value problem specified by Eqs.~(\ref{eq9}) and~(\ref{eq14}) for $L=10$
with slightly different initial conditions
\begin{eqnarray}
&&
\phi_{1,2}(\xi,\tau=0)=A_{1,2}\cos(\pi \xi/10),\nonumber\\
&&
\partial_{\tau}\phi_{1,2}(\xi,\tau=0)=0,\label{eq24}
\end{eqnarray}
where $A_{1}=5$ and $A_{2}=5.01$.
As is seen in Fig.\,7, a tiny difference in the initial conditions
leads to a quite different time evolution. This test assumes the existence
of chaotic dynamics and confirms the nonintegrability of the problem.
Consequently, finding exact analytical solutions is a hopeless task.

\begin{figure}[h]
\includegraphics{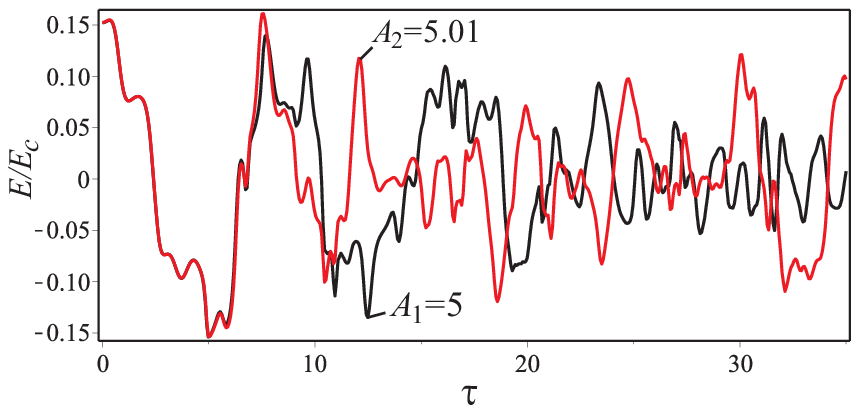}
\caption{Results of numerical integration of Eq.~(\ref{eq9}) with
the boundary conditions~(\ref{eq14}) for the initial conditions~(\ref{eq24}).
The black and red curves correspond to the normalized field oscillograms at $\xi=3$ for
the solutions with $A_{1}=5$ and $A_{2}=5.01$, respectively.}
\end{figure}

It should be emphasized that all the periodic solutions have been checked and an excellent agreement has been found between the pseudospectral (iterative) method and the calculations via MAPLE pdsolve in which a centered implicit scheme is employed.
Under the given initial
conditions, MAPLE yields the curves which are visually indistinguishable
from those of the pseudospectral method on all the plots in Figs.\,2 and 4--6. We
have also analyzed the stability of periodic solutions. Namely,
we have numerically integrated Eq.~(\ref{eq9}) with
the boundary conditions~(\ref{eq14}) for the initial conditions which
result from the mode shapes at $\tau=0$ and are augmented by small perturbations.
This analysis, performed for a number of situations, furnishes
a positive test for the overall robustness of the found standing wave solutions.
The exceptions are the resonant modes whose parameters $\Omega$ and $L$
correspond to small neighborhoods of the spikes in Fig.\,3. It is clear
that small perturbations lead to destruction of periodic solutions
in these cases. Thus, although the numerical tests are not a rigorous
mathematical proof of stability, we can state that the nonsingular
points ($dE_{m}/d\omega\neq\infty$) of the characteristics for $L\le 1.5$
in Fig.\,3 refer to the classically stable solutions.
In quantum theory, however, one can expect the decay of the bound
states which are related to the classical solutions corresponding
to the dotted branches in Fig.\,3. This is because
the existence of
such solutions is due to the false vacua of the potential [see Fig.\,1(a)].
Our numerical computations also show that the periodic solutions which
correspond to the true minimum of the potential $V$ [solid curves in Fig.\,3]
minimize the action, while the solutions labeled by the dotted curves correspond
to the local maxima of the action functional.

The classical time-periodic solutions obtained above can be quantized via
WKB methods~\cite{Das1,Das2,Jac}. The Bohr--Sommerfeld quantization
condition is written as
\begin{equation}
\int^{\pi/\Omega}_{0}\int^{L/2}_{-L/2} (\partial_{\tau}\phi)^2 d\xi
d\tau=k\pi,\label{eq25}
\end{equation}
where $k$ is a positive integer. Inserting Fourier representation~(\ref{eq20})
into Eq.~(\ref{eq25}), one obtains
\begin{equation}
\Omega L
\sum\limits^{M}_{m=1}\sum\limits^{N}_{n=1}
(2n-1)^{2}(\phi_{mn})^{2}=4k.\label{eq26}
\end{equation}
The coefficients $\phi_{mn}$ in Eq.~(\ref{eq26}) depend implicitly on
$\Omega$ and $L$. Hence, for fixed $L$ the implicit equation~(\ref{eq26})
defines the masses of quantum states. The corresponding values of
$\Omega$ for which Eq.~(\ref{eq26}) is satisfied with $k=1$ are shown in Fig.\,3
by the circles on the solid branches.
It is seen in Fig.\,3 that the lowest masses (frequencies) and
the related field strengths of quantum quasi-particles (plasmons), which are obtained from the quantization of
the stable periodic solutions, are about $\omega\sim 0.5\,m$ and $E\sim 0.05\,E_{c}$, respectively.
It is well known that in the general case, the
Bohr--Sommerfeld condition is valid only for large $k$ ($k\gg 1$).
Using the spatial periodicity of the found solutions, we can increase
the length of the
interval $|\xi|\le L/2$ by an integer number of half periods $L$. Thus, with
the replacement $L\to jL$, where $j$ is an arbitrary large integer,
Eq.~(\ref{eq26}) will correctly define the mass spectrum of the bound
states for $k\gg 1$. In classical theory, these states correspond to
the periodic solutions (nonlinear normal modes~\cite{Pet1,Pet2}) having
a large number of field variations on the interval of length $jL$.

\section{DISCUSSION}

Feynman shows in his seminal work~\cite{Fey1}, that the longitudinal waves
for which $A_{\sigma}=\partial_{\sigma}\varphi$ are nonphysical.
Indeed, the four-curl for these waves vanishes
($F_{\rho\sigma}=\partial_{\rho}\partial_{\sigma}\varphi-
\partial_{\sigma}\partial_{\rho}\varphi=0$)
and, accordingly, such a potential has no effect on a Dirac electron
since the transformation $\psi'=\exp(ig\varphi)\psi$,
in the notation corresponding to our (1+1)-dimensional case,
removes it. In our (1+1)-dimensional problem, we deal with the solenoidal
divergence-free vector potential $A^{\mu}=\varepsilon^{\mu\nu}\partial_{\nu}u$.
This field interacts with the spinor field and leads to the nonzero
component $E_{x}(x,t)$ of the longitudinal electric field oscillations.
Hence, our results do not contradict Feynman's conjecture.

It follows from the performed analysis that periodic oscillations of
the electric field, caused by oscillations of vacuum's charge polarization density
(collective excitations or plasmons of QED vacuum), can actually appear as
solutions of the homogeneous (1+1)-dimensional QED equations without
external sources. Probably, the most fundamental and complex question
on the subject is what sources can excite such oscillations in real world.
In what follows, we perform some qualitative analysis which gives evidence in favor of
coupling the longitudinal oscillations and the transverse electromagnetic
fields (photons) when the dispersive effects of polarized vacuum are taken into
account. It should be emphasized that any effects of this kind can be observed,
of course, only at very high field strengths that are expected to be created
by powerful X-ray lasers.

Let us consider, for example, (3+1)-dimensional QED fields in the region
between two perfectly conducting infinite plates located at $x=\pm l/2$ (Fabry--Perot
resonator). For the low photon energy and a weak field, i.e.,
\begin{equation}
\omega\ll m,\quad |{\bf E}|\ll E_{c},\quad \omega/m\ll |{\bf E}|/E_{c},\label{eq27}
\end{equation}
one can employ the effective field theory approach represented by the
Heisenberg--Euler Lagrangian with small dispersive corrections~\cite{Roz,Gri,Gus,Dit,Dun,Mar,Lun}.
For simplicity, we will assume that the nonzero electromagnetic field
components are $E_{x}(x,t)$, $E_{y}(x,t)$, and $B_{z}(x,t)$, for which
${\bf E}\cdot {\bf B}=0$.
Vacuum polarization effects can be taken into account using the following
modification of the constitutive relations in Maxwell's theory:
\begin{eqnarray}
{\bf D}={\bf E}+{\hat F}(S,\square){\bf E},\nonumber\\
{\bf H}={\bf B}+{\hat G}(S,\square){\bf B},\label{eq28}
\end{eqnarray}
where $S={\bf E}^2-{\bf B}^2$.
If conditions~(\ref{eq27}) hold, one obtains~\cite{Mar,Lun}
\begin{equation}
{\hat F}={\hat G}={8\over 45}{\alpha^{2}\over m^{4}}S-{4\over 15}
{\alpha\over m^{2}}\square.\label{eq29}
\end{equation}
Suppose now that $\omega$ approaches $m$. Then inequalities~(\ref{eq27})
are not satisfied and approximation~(\ref{eq29}) is not applicable.
Nevertheless, due to the correspondence principle, coherent states of
an electromagnetic field will still be described by the Maxwell equations with
the constitutive relations~(\ref{eq28}). An explicit form of
the operators ${\hat F}$ and ${\hat G}$ in this case is unknown at present.
To ensure Lorentz invariance and the fundamental property that the
traveling plane waves for which ${\bf E}^{2}={\bf B}^{2}$ and $\square {\bf E}=0$
do not polarize vacuum, these operators should depend
only on the invariant $S$ and the wave operator~$\square$. Hence, the equations
for nonzero field components can be written as
\begin{eqnarray}
&&\partial_{x}H_{z}=-\partial_{t}D_{y},\label{eq30}\\
&&\partial_{x}E_{y}=-\partial_{t}B_{z},\label{eq31}\\
&&\partial_{x}D_{x}=0.\label{eq32}
\end{eqnarray}
It follows from Eqs.~(\ref{eq30})--(\ref{eq32}) that there exist degenerate nontrivial
solutions for which $E_{x}\neq 0$, while all the other field components vanish and
\begin{equation}
D_{x}=E_{x}+{\hat F}(E^{2}_{x},\square)E_{x}=0.\label{eq33}
\end{equation}

There is a reason to believe that the operator ${\hat F}$ should
be such that Eq.~(\ref{eq33}) will be equivalent
or nearly equivalent to the MSG equation~(\ref{eq9}).
This assumption is partially supported by the results of a recent work~\cite{Lv}.
The authors of that work have examined the accuracy of an
intrinsically (1+1)-dimensional QED to predict
the forces and charges of a three-dimensional system that has
a high degree of symmetry and therefore depends effectively only
on a single spatial coordinate.
There are no transverse electromagnetic waves or photons in
(1+1)-dimensional QED (massive Schwinger model). However, as it has been
shown in~\cite{Lv}, this model is not merely ``a game of mind'' and can
correctly predict the vacuum polarization in a flat capacitor. The
static electric displacements in (1+1)-dimensional and (3+1)-dimensional theories
have been found to differ only slightly. Consequently, our assumption
that the constitutive relations $D_{x}(E_{x})$ [see Eq.~(\ref{eq33})] are
nearly equivalent in both theories is supported by the results of~\cite{Lv}
in the static limit. Since the (3+1)-dimensional theory admits the
time-dependent solutions in which there are no transverse electromagnetic
fields and only the longitudinal electric field exists, it is natural to assume
that these solutions can be described by an intrinsically (1+1)-dimensional QED,
i.e., the corresponding time-dependent electric displacements are also
nearly equivalent in both theories. To mathematically prove
this assumption, one needs to find a nonperturbative explicit
expression for the operator $\hat F$. This problem still remains
unsolved at present.

In a more general case where the transverse (with respect to the $x$ direction)
field components $E_y$ and $B_z$ are nonzero, it follows from Eqs.~(\ref{eq30})--(\ref{eq31})
that the equations for these components are coupled with Eq.~(\ref{eq32}).
The coupling is due to the
nonlinearity of relations~(\ref{eq28}), in which $S=E^{2}_{x}+E^{2}_{y}-B^{2}_{z}$.
Thus, longitudinal oscillations can be excited by the transverse standing
waves in the Fabry--Perot resonator at the frequencies $\omega\lesssim m$.
The presence of longitudinal modes (plasmons in quantum theory) will affect
counter-propagating photons and, hence, these modes cannot be ignored as nonphysical.
It should also be noted that the magnetic-field component in such a hybrid
longitudinal-transverse field configuration will significantly reduce
the pair creation rate compared to the purely electric field case~\cite{Ruf}.

\section{CONCLUSIONS}

In this article, we have studied the problem of plasmons in QED vacuum. It
has been shown that the bosonized version of (1+1)-dimensional QED admits
the existence of classical stable time-periodic solutions, i.e., standing
waves of the longitudinal electric field and vacuum's polarization density.
We have numerically calculated mode shapes and field oscillograms for
these solutions. The region of the existence of solutions in the parameter
space has been established and numerical tests of their robustness
have been performed. Applying the Bohr--Sommerfeld quantization condition,
we have determined the mass spectrum of charge-zero bound states (plasmons)
which correspond in quantum theory to the found classical solutions.
We have also presented qualitative analysis which gives evidence in favor
of coupling the longitudinal oscillations (plasmons) and the transverse
waves (photons) in the dispersive vacuum. The performed analysis
predicts the appearance of plasmons of QED vacuum in colliding
laser pulses at the frequencies $\omega\sim 0.5\,m$.
The required strength of the incoming laser field can be estimated as
$E\geq 0.05\,E_{c}$.
This can be relevant for future experiments with powerful X-ray and
gamma lasers. To obtain a more precise estimate of the field strength, one needs to solve the (3+1)-dimensional problem with coupling between the longitudinal and transverse modes. However, the solution of this extremely hard problem falls beyond the scope of our article.

\acknowledgements

This work was supported by the Russian Science Foundation
(Project No.~14--12--00510, Secs. I--III)
and the Russian Government
(Contract No.~14.B25.31.0008, Sec. IV).

\bibliography{Petrov_Kudrin_bibl}

\end{document}